\documentclass[aps,twocolumn]{revtex4}
\usepackage{graphicx}

\newcommand{\be}{\begin{equation}}
\newcommand{\ee}{\end{equation}}
\newcommand{\bqn}{\begin{eqnarray}}
\newcommand{\eqn}{\end{eqnarray}}

\begin{document}

\title{Vacuum polarization for compactified  $QED_{4+1}$ in a
magnetic flux background}

\author{C. Ccapa Ttira and C.D. Fosco}
\affiliation{Centro At\'omico Bariloche and Instituto Balseiro, 
 Comisi\'on Nacional de Energ\'\i a At\'omica,  
8400 Bariloche, Argentina}
\author{A. P. C. Malbouisson and I. Roditi}
\affiliation{Centro
Brasileiro de Pesquisas F\'{\i}sicas/MCT, 22290-180, Rio de
Janeiro, RJ, Brazil}

\begin{abstract}
We evaluate one-loop effects for $QED_{4+1}$ compactified to
${\bf R}^4 \times S^1$, in a non-trivial vacuum for the gauge
field, such that a non-vanishing magnetic flux is encircled along the extra
dimension. We obtain the vacuum polarization tensor and evaluate the exact
parity breaking term, presenting the results from the point of view of the
effective $3+1$ dimensional theory.
\end{abstract}

%\date{\today}

\maketitle

\section{Introduction}\label{sec:intro}

Quantum field theory models with compactified dimensions have been used
to describe different physical situations, ranging from finite
size effects in critical phenomena~\cite{Brezin:1985xx,zinn,livro}, to  the
unification of fundamental interactions~\cite{dvali,sundrum,antoniadis1}. 
These ideas have recently attracted renewed interest; for instance,
results about the electroweak phase transition has been
presented in~\cite{Panico}, in the context of a $4+1$-dimensional theory
with a compactified dimension. 

The presence of extra compactified dimensions may give rise to effects at 
different scales, not just those in high energy physics realm;  in particular
on low energy phenomena, like atomic physics. Following this idea, we evaluate,
within a specific  domain of physics, namely Quantum Electrodynamics, the
effects that would follow if our world were 5-dimensional. In particular, we
investigate at one-loop order some effects due to the assumption of a
non-trivial vacuum with a non-vanishing magnetic flux along the compactified
dimension. 

Effects on the anomalous magnetic moment of the muon associated with
extra-dimensional excitations of the photon and of the W and Z bosons, have 
been studied in~\cite{Nath}, in a space whose extra dimensions have large
compactification radii. Those authors 
have shown that, when the extra-dimensional corrections to the Fermi constant
are
included, their  effects on $(g_{\mu}-2)$ become too small to be observable.
They discuss a model which avoids the extra-dimensional corrections to the
muon 
decay $\mu\rightarrow e{\bar{\nu}}_e\nu_{\mu}$ without suppression of their
effects on $(g_{\mu}-2)$. Eventual extra-dimensions effects on $(g_{\mu}-2)$
would be very interesting. We know, since the  $g-2$ experiment at Brookhaven
National Laboratory (USA) in 2004, and subsequent experiments, that the
expected value from standard theoretical calculations, that predicts $g=2$,
could not be confirmed, since both theoretical prediction and experimental
results have a large amount of uncertainty. Although a 
conclusive response is not available, a value of $g\neq 2$ is not
excluded by experimentalists~\cite{babar}. Indeed, in the experimental framework
of $QED$ a recent experiment for the electron magnetic moment, gives a much more
precise value for $g_e$ (the claimed uncertainty is nearly 6 times lower than in
the past). These authors still find a deviation from the value
$g=2$~\cite{Harvard}. In atomic physics, very accurate measurements of the
asymptotic quantum effects on Rydberg excitations have also been carried
out~\cite{Deuterium}.   
   
Another interesting consequence of the possible existence of extra dimensions is
explored in~\cite{Dubovsky}. This study shows that they would imply that
electric charge might not be exactly conserved, what has been a subject of
discussion for a long
time~\cite{Ignatev,Mohapatra,Susuki}. 
As mentioned in~\cite{Dubovsky}, in four-dimensional theories, a tiny deviation
from electric charge conservation would lead to contradictions with low-energy
tests
of $QED$. These could, in turn, be cured by the introduction of 
hyphotetical millicharged particles~\cite{Susuki}. However, as argued
in~\cite{Dubovsky},  if our world were considered as 
 a submanifold of a higher-dimensional space, this artifact would not be
necessary. Indeed 
 in this 
case, particles initially confined to our $4$-dimensional subspace could, under
some circumstances 
migrate to the extra dimensions. 
The idea presented in~\cite{Dubovsky} is that if they are electrically charged, their migration from our world 
into extra dimensions would appear for us as non-conservation of electric
charge. 
Charge non-conservation and other possible effects of extra dimensions could
perhaps be investigated in experiments similar to those
in~\cite{Harvard,Deuterium}.

In~\cite{gregos} a $U(1)$ gauge field theory with fermion or scalar fields
defined on a space with extra compactified dimensions has been considered. These
authors compute the fermion-induced quantum energy in the presence of a constant
magnetic field directed towards the $z$-axis. They study the effect of extra
dimensions on the asymptotic behavior of the quantum energy in the strong field
limit, and find that the weak logarithmic growth of the quantum energy in four
dimensions is modified by a rapid power growth in a space-time with extra
dimensions.

%An analysis of the  The Kaluza-Klein effects on g-2 are found to be very sensitive to the %number of extra dimensions. For models where the quark-lepton generations live on the 
%four-dimensional wall, it is  A model  is discussed 

Because of the reasons described above, we believe that the study of effects due
to extra dimensions in electromagnetic phenomena is a subject of actual
interest. We present here new results about that topic; they correspond to
quantum effects in $QED$ with an extra dimension, in a magnetic flux
background. In particular, we consider the modifications that the extra
dimension produces on the vacuum polarization phenomenon. 

This paper is organized as follows: in section~\ref{sec:themodel} we
introduce the model and study its more important features, mostly related
to the realization of gauge invariance within the context of a theory with
a compactified dimension, having an extra dimensional-like mode expansion in
mind.  Section~\ref{sec:effective} deals with the effective action, which
is derived including both parity-conserving and parity-violating parts.  In~\ref{sec:nonparity}, we apply the general results
of~\ref{sec:effective} to the exact calculation of the parity conserving
part of the vacuum polarization tensor. 
In~\ref{sec:parity}, we consider parity-breaking
effects. Section~\ref{sec:conc} contains our conclusions.

\section{Theoretical framework}\label{sec:themodel}

From a general point of view one can consider a simply or non simply-connected $D$-dimensional
manifold with a topology ${\bf R}^{D}_{d}={\bf R}^{D-d}\times
S_{l_{1}}\times {\mathcal S}_{l_{2}}\cdots \times {\mathcal S}_{l_{d}}$,
with $l_{1}$ corresponding to the inverse temperature and $l_{2}\,,\cdots
l_{d}$  to the compactification of $d-1$ spatial
dimensions (this case has been considered, within the context of spontaneous 
symmetry breaking, in~\cite{Ademir}). 
An interesting yet simple example of this, corresponds to the
compactification of one dimension in an ${\bf R}^D$ Euclidean spacetime, such
that the topology of the resulting manifold ${\mathcal M}$ is that of
\mbox{${\bf R}^{D-1} \times {\mathcal S}_ 1$}, i.e., `circular
compactification'.  Although the compelling features that emerge in this
situation have been studied using several different techniques in the
literature, one can take advantage of a (formal) common feature; indeed,
they share many properties with the imaginary-time formulation of quantum
field theory at finite temperature~\cite{3mats1,3ume4}.  This allows one,
for example, to take advantage of the many well-known methods and results
developed in this context, such as Feynman diagrams and renormalization
techniques, to import them to the case under consideration. 

For just one compactified dimension (imaginary time or a spatial
dimension) the Feynman rules are modified, the most characteristic new
feature is the Matsubara prescription for momentum integrals,  
\begin{equation}
\int \frac{dk_s}{2\pi }\rightarrow \frac 1{\xi}\sum_{n=-\infty }^{+\infty
}\;;\;\;\;k_s\rightarrow \frac{2n\pi }{\xi}\;,
\label{Matsubara1}
\end{equation}
where $k_s$ amounts to the momentum component corresponding to the compactified
dimension, while $\xi$ equals $\beta$ or $L$, for the finite temperature and
compactified spatial dimension cases, respectively.

Within the previous general framework, we here investigate one-loop effects for
$QED_{3+1}$ with an extra compactified dimension, in a non-trivial vacuum for
the gauge field, defined by a non-vanishing component along the extra dimension.

The system we shall deal with may be conveniently defined in terms of an 
Euclidean action, $S$, which has the structure:
\begin{equation}\label{eq:defs0}
S({\mathcal A};{\bar\Psi},\Psi) \;=\; S_g({\mathcal A}) 
\,+\, S_f({\mathcal A};{\bar\Psi},\Psi)\;, 
\end{equation}
where $S_g$ and $S_f$ denote the $U(1)$ gauge field and fermionic actions,
respectively. The former is assumed to have a standard Maxwell form, namely:
\begin{equation}\label{eq:defsg}
S_g({\mathcal A}) \;=\; \frac{1}{4} \, \int d^5x \, {\mathcal
F}_{\alpha\beta}  {\mathcal F}_{\alpha\beta} \;,
\end{equation}
with ${\mathcal F}_{\alpha\beta}\equiv\partial_\alpha {\mathcal A}_\beta -
\partial_\beta {\mathcal A}_\alpha$, where we adopted the convention that
indices from the beginning of the Greek alphabet ($\alpha$, $\beta$,
\ldots) label all the coordinates of the spacetime manifold, and therefore
run from $0$ to $4$.  Since we will be specially interested in the model as
it is seen from a $3+1$ dimensional point of view, we shall also use
another convention: indices from the middle of the Greek alphabet ($\mu$,
$\nu$, \ldots) are reserved for the $3+1$-dimensional spacetime
coordinates while, when this notation is used, the extra dimension
coordinate shall be denoted by $s$. Then:
\begin{eqnarray}
\alpha \,&=&\, 0,\,1,\,2,\,3,\,4 \;,\;\;\;
\mu \,=\, 0,\,1,\,2,\,3\;,\;\;\;\nonumber \\ 
&& d^5 x \;\equiv\; d^{3+1}x \,dx_4 \;=\; d^{3+1}x \, ds \;,
\end{eqnarray}
and $x$ will be assumed to denote the $3+1$ coordinates $x_\mu$,
unless explicit indication on the contrary.  The extra dimension is assumed
to be compactified with a radius $R$, so that $s \sim s + L$, $L = 2 \pi
R$.

On the other hand, the Dirac action, $S_f$, is given by
\begin{equation}
 S_f({\bar\Psi},\Psi;{\mathcal A}) \;=\; \int d^{3+1}x \, ds \; {\bar\Psi}(x,s) 
\big( {\mathcal D} + m \big) \Psi(x,s)
\end{equation} 
where ${\mathcal D}$ is the $4+1$ dimensional Dirac operator, \mbox{${\mathcal
D}= \gamma_\alpha D_\alpha$}. The covariant derivative 
\mbox{$D_\alpha \equiv \partial_\alpha + i g {\mathcal A}_\alpha$} 
includes a coupling constant $g$ with the dimensions of $({\rm
mass})^{-\frac{1}{2}}$. For Dirac's $\gamma$-matrices, we assume that
$\gamma_s \equiv \gamma_5$, where the latter is the $\gamma_5$ matrix for
the $3+1$ world. 

To proceed, we discuss now the mode expansion and its relation to gauge
invariance. To that end, we follow~\cite{zinn}, where this issue is
discussed at length, albeit in the finite temperature theory context, in the
Matsubara formulation of thermal field theory. Due to the formal analogy  
with this situation, a quite straightforward procedure allows us to adapt 
the results derived there to our case. The necessary changes that 
follow from the fact that our compactified dimension is spatial rather 
than temporal are taken into account by using (\ref{Matsubara1}). 
In that analogy, the length $L$ plays the same role of 
the inverse temperature in \cite{zinn}: $L \sim \beta$, $\beta = T^{-1}$.

What follows is a brief review of some of those properties (the ones which
are relevant to our study), adapted to our case and conventions. To begin
with, the gauge field configuration ${\mathcal A}_\alpha(x,s)$ may be
decomposed into its zero ($A_\alpha$) and non-zero ($Q_\alpha$) mode
components:
\begin{equation}
{\mathcal A}_\alpha(x,s) \;=\; L^{-\frac{1}{2}} \, A_\alpha(x) \,+\,
Q_\alpha(x,s) \;,
\end{equation} 
where the two terms in this decomposition may be defined by:
\begin{equation}
A_\alpha(x) \;=\;  L^{-\frac{1}{2}} \, \int_0^L ds \, 
{\mathcal A}_\alpha (x,s) \;,
\end{equation}
and 
\begin{equation}
Q_\alpha(x,s) \;=\; {\mathcal A}_\alpha(x,s) \,-\, L^{-\frac{1}{2}} \,
A_\alpha(x) \;, 
\end{equation} 
so that $\int_0^L ds \, Q_\alpha(x,s)\,=\, 0$. An $L^{-\frac{1}{2}}$ factor
has been included  in the zero mode term in order to make this field have
the usual mass dimensions in $3+1$ spacetime dimensions; this property will
become useful after dimensional reduction. 

The decomposition above finds a natural interpretation when one considers the
Fourier expansion of the gauge field along the extra dimension:
\begin{equation}
{\mathcal A}_\alpha (x,s) \;=\; L^{-\frac{1}{2}} \, 
\sum_{n=-\infty}^{\infty}\, e^{ i \omega_n s} \,\widetilde{\mathcal A}_\alpha(x,n) \;,
\end{equation}
with $\omega_n \equiv \frac{2 \pi n}{L}$, where one identifies:
\begin{equation}
A_\alpha(x) \,=\, \widetilde{\mathcal A}_\alpha(x,0) \;\;,\;\;\;\; 
Q_\alpha(x,s) \,=\, L^{-\frac{1}{2}} \, 
\sum_{n \neq 0}\, e^{ i \omega_n s} \,\widetilde{\mathcal A}_\alpha(x,n) \;.
\end{equation}

Then we dimensionally reduce the theory, what, for the gauge field action,
amounts to keeping just the zero mode component of the gauge field. Thus:
\begin{equation}
S_g({\mathcal A}) \; \to \; S_g (A) \;=\; S_g(A_\mu, \,A_s) \;, 
\end{equation}
where:
\begin{equation}
S_g(A_\mu, A_s)=\int d^{3+1}x \, \Big[
\frac{1}{2} \partial_\mu A_s \partial_\mu A_s 
+\frac{1}{4} F_{\mu\nu}(A) F_{\mu\nu}(A) \Big],
\end{equation}
with $F_{\mu\nu}(A) \equiv \partial_\nu A_\nu - \partial_\nu A_\mu$.

Regarding the fermionic action $S_f$, the reduction amounts to:
\begin{equation}
S_f({\mathcal A}; {\bar\Psi}, \Psi ) \; \to \; 
S_f(A_\mu,A_s;{\bar\Psi},\Psi) \;. 
\end{equation}
The fermionic field is not dimensionally reduced by the simple reason that,
in the calculation of the effective gauge field action, its only contribution
comes from the fermion loop. That loop may be represented as a series of
$3+1$ loops, each one with a different mass. Although the contributions of
heavier modes may be relatively suppressed, the very fact that there is an
infinite number of them forbids us to truncate that series (even if there
were a zero mode).

Thus, the following explicit expression for the fermionic action shall be
used after dimensional reduction:
\begin{equation}
S_f \;=\; \int d^{3+1} x \, \int_0^L ds \; {\bar\Psi}(x,s) 
\big( \not \!\! D + \gamma_s D_s  + m \big)  \Psi(x,s)
\end{equation} 
where
\begin{equation}
\not \!\! D \,=\, \gamma_\mu (\partial_\mu + i e A_\mu )
\;\; D_s \,=\, \partial_s + i e A_s  \;.
\end{equation}
We have introduced a new, dimensionless coupling constant $e \equiv g
L^{-\frac{1}{2}}$, which shall play the role of the electric charge in $3+1$
dimensions.

As explained in~\cite{zinn}, when considering the form of the gauge
transformations in terms of the decomposition into zero and non-zero modes,
one finds that it $A_\mu$ transforms as a standard gauge field (in $3+1$ dimensions):
\begin{equation}
\delta A_\mu (x) \,=\, \partial_\mu \alpha (x)
\end{equation}
while its extra dimensional component $A_s$, a scalar from the
$3+1$-dimensional point of view, is shifted by a constant:
\begin{equation}
\delta A_s (x) \,=\,\Omega \;.
\end{equation}
The constant $\Omega$ has to be of the form~\mbox{$\Omega = \frac{2 \pi
k}{L e}$}, where $k$ is an integer, since the gauge field is
coupled to a (charged) fermionic field, whose transformation law under
simultaneous action of the previous gauge transformations is:
\begin{eqnarray}
\Psi (x,s) &\to&  e^{-i e [\alpha(x) + \Omega s ]} \, \Psi (x,s) \nonumber\\
\bar{\Psi} (x,s) &\to&  e^{ i e [\alpha(x) + \Omega s ]} \,
\bar{\Psi} (x,s) \;.
\end{eqnarray}

\section{Effective action}\label{sec:effective}  

We now define the part of the effective action that only
depends on the (dimensionally reduced) gauge field, $\Gamma (A)$, 
\begin{equation}
\Gamma (A) \; \equiv \; \Gamma (A; {\bar\Psi}, \Psi) \Big|_{{\bar\Psi}=
\Psi = 0} \;,
\end{equation}
where $\Gamma (A; {\bar\Psi}, \Psi)$ is the full effective action. 
The functional $\Gamma(A)$ allows one to derive $1PI$ functions containing 
only $A_\mu$, $A_s$ external lines. The former have an immediate $3+1$
dimensional interpretation, while the latter shall be assumed to have
a constant (but otherwise arbitrary) value, which is determined by a
condition which is external to the model.

On the other hand, at the one-loop order, the only non-trivial term comes 
from the fermionic loop:
\begin{equation}
\Gamma(A) \;=\; \Gamma^{(0)}(A) \,+\,\Gamma^{(1)}(A) \;+\;\ldots  
\end{equation}
where $\Gamma^{(0)}(A) = S_g(A)$ and 
\begin{equation}
e^{-\Gamma^{(1)}(A)} \;=\; \int {\mathcal D}\Psi {\mathcal D}{\bar\Psi} e^{-
S_f(A; {\bar\Psi},\Psi) } \;. 
\end{equation}

We shall focus on the effective action for the gauge field components
$A_\mu$ that have a direct physical interpretation from a $3+1$-dimensional
perspective. Regarding the scalar component, $A_s$, as we have said above, it 
will be assumed to yield a non-vanishing flux:
\begin{equation}
e \int_0^L ds A_s \;=\; \theta 
\end{equation}
where $\theta$ is a constant.  This condition may be conveniently solved by
means of a constant $A_s$: 
\begin{equation}
A_s \;=\; \frac{\theta}{e L} \;,
\end{equation}
which is the gauge fixing that we shall assume. Note that, since the gauge
transformations shift $A_s$ by an integer multiple of $\frac{2\pi}{e L}$,
we may fix the value of $\theta$ to the fundamental region:
\begin{equation}
0 \leq 	\theta < 2 \pi  \;,
\end{equation}
which we shall assume in what follows.

It is worth noting that this kind of gauge field configuration may be
interpreted as `topological', in the sense that it corresponds locally
(althougth not globally) to a `pure gauge' field configuration. Indeed, it
cannot be gauged away, since the corresponding gauge transformation would
be multivalued (when the extra dimension is encircled). Charged fields feel
this kind of configuration when they encircle the extra coordinate, in a fashion
that resembles the Aharonov-Bohm effect. The field configuration may be realized
in a similar way to this effect: a singular field strength pointing in a
direction orthogonal to the plane of the circle. Besides, as in the
Aharonov-Bohm effect, the region of space where the field strength is
non-vanishing, cannot be reached by the charged fields. The situation can be
easily visualized in a lower dimensional example, namely, the case of a $2+1$
dimensional theory, if one assumes $x_2$ to be the extra, compactified
dimension. Here, space is a cylinder, and the gauge field configuration
corresponding to the vacuum field would be a singular flux string along the
cylinder axis. This means that it is outside of the assumed cylindrical space,
since it needs a third coordinate to be realized. In a similar way, the kind of
configuration we consider could be realized by singular, monopole-like field
strengths in a higher (more than $4+1$) dimensional manifold.

We then proceed to Fourier expand the fermionic fields along the $s$
coordinate:
\begin{eqnarray}
\Psi(x,s) &=& L^{-\frac{1}{2}} \,\sum_{n=-\infty}^\infty e^{ i \omega_n s} \psi_n(x)
\nonumber\\
{\bar\Psi}(x,s) &=& L^{-\frac{1}{2}} \,\sum_{n=-\infty}^\infty e^{ - i \omega_n s}
{\bar\psi}_n(x) \;,
\end{eqnarray}
and insert this into the functional expression for $\Gamma^{(1)}(A)$, to
obtain:
\begin{equation}\label{eq:sfexpand}
S_f \;=\; \sum_{n=-\infty}^{n=+\infty} 
\int d^{3+1}x\,\bar{\psi}_n(x) \big(\not \!\! D + i \gamma_s (\omega_n +
\frac{\theta}{L} ) + m \big) \psi_n(x) \;.
\end{equation}
Under the same expansion, the fermionic measure factorizes:
\begin{equation}
\mathcal{D}\Psi \mathcal{D}\bar{\Psi}\;=\;
\prod_{n=-\infty}^{n=+\infty} \mathcal{D}\psi_n(x) \mathcal{D}\bar{\psi}_n(x),
\end{equation}
and, finally, the Euclidean action corresponding to each mode $n$ may be
equivalently written as follows
$$
\int d^{3+1}x\,\bar{\psi}_n(x) \big(\not \!\! D + i \gamma_s (\omega_n +
\frac{\theta}{L} ) + m \big) \psi_n(x) 
$$
\begin{equation}
=\; \int d^{3+1}x \,\,\bar{\psi}_n(x)(\not \!\! D + M_n \,
e^{-i \varphi_n \gamma_5})\psi_n(x)
\end{equation}
with
\begin{equation}
 M_n \equiv \sqrt{m^2+(\omega_n+\theta/L)^2}\;,\;\;\; 
\varphi_n = {\rm arctan}(\frac{\omega_n+\theta/L}{m}) \;.
\end{equation}
The existence of a $\gamma_5$ term means that parity symmetry will generally be
broken; to study that phenomenon more clearly, we perform  a 
change in the fermionic variables that gets rid of the dependence in $\gamma_5$,
\begin{equation}
\psi_n(x)\to e^{-i\gamma_5 \varphi_n/2}\psi_n(x)\,,\,\,\,\,\bar{\psi}_n(x) 
\to \bar{\psi}_n(x)e^{-i\gamma_5 \varphi_n/2} \;,
\end{equation}
after which the mode labelled by $n$ has the action:
\begin{equation}
\int d^{3+1}x \,\,\bar{\psi}_n(x)(\not \!\! D + M_n \,
e^{-i \varphi_n \gamma_5})\psi_n(x) \;.
\end{equation}
This chiral rotation in the $3+1$ Euclidean fermionic variables induces,
however an anomalous Jacobian ${\mathcal J}_n$ for each mode. 
Then, $\Gamma^{(1)}$ may be written as follows:
\begin{equation}
e^{-\Gamma^{(1)}(A)} \;=\; \prod_{n=-\infty}^{+\infty}\Big[  
{\mathcal J}_n \; e^{-\Gamma^{(1)}_{3+1}(A,M_n)} \Big]\;,
\end{equation}
where 
\begin{equation}\label{eq:defj}
{\mathcal J}_n \;=\; \exp \big(\frac{i e^2}{16 \pi^2} \, 
\varphi_n \int d^{3+1}x {\tilde F}_{\mu\nu} F_{\mu\nu} \big)  \;,
\end{equation}
with ${\tilde F}_{\mu\nu}=\frac{1}{2} \epsilon_{\mu\nu\rho\lambda}
F_{\rho\lambda}$,
and $\Gamma^{(1)}_{3+1}(A,M_n)$ is the one-loop fermionic contribution to the
effective action, for a fermion whose mass is $M_n$, in $3+1$ dimensions. 
Of course, it may be expressed as a fermionic determinant:
\begin{equation}
e^{-\Gamma^{(1)}_{3+1}(A,M_n)} \;=\; \det(\not \!\! D + M_n) \;.
\end{equation}
Then, we arrive to a general expression for the one loop effective action,
\begin{equation}
\Gamma^{(1)}(A) \;=\;\Gamma^{(1)}_e (A) \,+\, \Gamma^{(1)}_o (A) 
\end{equation}
where the $e$ and $o$ subscripts stand for the even an odd components
(regarding parity transformations) and are given by
\begin{equation}
\Gamma^{(1)}_e(A) \;=\; \sum_{n=-\infty}^{\infty} \Gamma^{(1)}_{3+1}(A,M_n) 
\end{equation}
and 
\begin{equation}\label{eq:sumj}
\Gamma^{(1)}_o(A) \;=\; - \sum_{n=-\infty}^{\infty} \ln {\mathcal J}_n \;,
\end{equation}
respectively.

\section{Parity conserving term}\label{sec:nonparity}

The parity conserving part of the effective action may be obtained by
performing the sum of the required $QED_{3+1}$ object, with an $n$-dependent
mass, $M_n$.  We shall focus on that part of $\Gamma^{(1)}_e$ that
contributes to the vacuum polarization tensor for the $A_\mu$ gauge field components.
Since we are not interested in response functions which involve the $s$
component of the currents, it is useful to define:
\begin{equation}
\Gamma^{(1)}_e (A_\mu) \;\equiv\;\Gamma^{(1)}_e (A_\mu,A_s) \,-\,
\Gamma^{(1)}_e (0,A_s) \;.   
\end{equation}
Note that $\Gamma^{(1)}_e (0,A_s)\equiv \Gamma_s(A_s)$ does not contribute
to response functions involving $A_\mu$, although it can be used to study
the fermion-loop corrections to an $A_s$ effective potential. The explicit
form of this function is~\cite{zinn}:
\begin{equation}
 \Gamma_s(A_s) \,=\,- 2 L \int d^{3+1}x \, \int \frac{d^4k}{(2\pi)^4} 
\ln \big[ \cosh(L k) \,+\, \cos \theta \big] \;. 
\end{equation}

The vacuum polarization tensor $\Pi_{\mu\nu}$  is obtained from the quadratic 
term in a functional expansion in the gauge field:
\begin{equation}
\Gamma^{(1)}_e(A_\mu) = \frac{1}{2} \int d^{3+1}x \int d^{3+1}y 
A_\mu(x) \Pi_{\mu \nu}(x,y) A_\nu(y) + \ldots 
\end{equation}

It is then sufficient to resort to the analogous expansion for the $3+1$ 
dimensional effective action,
\begin{eqnarray}
\Gamma_{3+1}^{(1)}(A,M_n)&=& 
\frac{1}{2} \int
d^{3+1}x \int d^{3+1}y \nonumber \\ 
&\times&\left[A_\mu(x) \Pi_{\mu \nu}^{(n)}(x,y) A_\nu(y)\right]
+\ldots 
\end{eqnarray}
(which is even) so that the vacuum polarization receives contributions from
all the modes:
\begin{equation}\label{eq:pisum}
\Pi^e_{\mu \nu} \;=\;\sum_n \Pi_{\mu \nu}^{(n)}\,,
\end{equation}
where $\Pi_{\mu \nu}^{(n)}=\Pi^{(n)}(k^2) \; \delta^T_{\mu \nu}(k)$, with:
\begin{equation}\label{eq:tamed}
\Pi^{(n)}(k^2) \;=\;
\frac{2\,e^2}{\pi} \int_0^1 d\beta\,\,\beta(1-\beta)\,\,\ln 
\left[1+\beta(1-\beta)\frac{k^2}{M_n^2}\right]\;,
\end{equation}
which is formally identical to the renormalized scalar part of the vacuum 
polarization tensor for a $3+1$ dimensional theory,
and the transverse projector is defined by: 
\mbox{$\delta^T_{\mu \nu}(k) \equiv  \delta_{\mu\nu} - k_\mu k_\nu/k^2$}.
Note that the renormalization performed for $\Pi^{(n)}(k^2)$ should in fact be interpreted
as a subtraction for the $4+1$ dimensional theory, which (see below) yields a logarithmically divergent
vacuum polarization, as in $3+1$ dimensions, once all the symmetries have been taken into account. 
The subtraction already performed in $3+1$ dimensions does not yet fulfill the
renormalization conditions for the $4+1$ dimensional theory: the zero of
$\Pi^{(n)}$ is at $k^2=0$ for each term, but the limit $k^2
\to 0$ does not necessarily commute with the (infinite) sum over modes. Indeed,
that commutativity is not guaranteed, since the series in (42) does not converge
uniformly.

To do have the proper pole in the propagator, we shall need to perform also a
finite renormalization. Indeed, the sum in (\ref{eq:pisum}) may be
explicitly evaluated using $zeta$-function regularization
techniques~\cite{EE}; we can write $\Pi(k^2)=\sum_{n}\Pi^{(n)}(k^2)$, with 
\begin{equation}\label{eq:tamed}
\Pi(k^2) \;=\;
\frac{2\,e^2}{\pi} \int_0^1 d\beta\,\,\beta(1-\beta)\Pi(k^2,\beta),
\end{equation}
and  
\begin{equation}\label{eq:tamed1}
\Pi(k^2,\beta) \;=\;
\sum_{n=-\infty}^{+\infty}\ln\left[\frac{(bn+\frac{\theta}{L})^2+m^2+\beta(1-\beta)k^2}{(bn+\frac{\theta}{L})^2+m^2}
\right]\end{equation}
where $b=2\pi/L$. Then it can be readily seen that,
\begin{eqnarray}
\Pi(k^2,\beta)&=&\lim_{s\rightarrow 0_{-}}\left[\frac{d}{ds}\left(Z_1^{m^2}(s,b^2,\frac{\theta}{L})\right)\right]\nonumber \\
&& -\lim_{s\rightarrow 0_{-}}\left[\frac{d}{ds}\left(Z_1^{m^2+\beta(1-\beta)k^2}(s,b^2,\frac{\theta}{L})\right)\right],\nonumber \\
\label{zeta1}
\end{eqnarray}
where $Z_1(s,...)$ are generalized inhomogeneous $zeta$-functions,
\begin{equation}
Z_1^{M^2}(s,b^2,\frac{\theta}{L})=\sum_{n=-\infty}^{\infty}\left[(bn+\frac{\theta}{L})^2+m^2\right]^{-s},
\label{zeta2}
\end{equation}
with $M^2= m^2$ or $M^2=m^2+\beta(1-\beta)k^2$;  

The generalized inhomogeneous $zeta$-function can be written in the whole complex $s$-plane in the form~\cite{EE}, 
\begin{widetext}
\begin{equation}
Z_1^{M^2}(s,b^2,\frac{\theta}{L})=\frac{\sqrt{\pi}}{\Gamma(s)}\left[\Gamma(s-\frac{1}{2})M^{1-2s}+ \sum_{n=1}^{\infty}\left(\frac{\pi n}{Mb}\right)^{s-\frac{1}{2}}cos(n\theta)K_{s-\frac{1}{2}}\left( \frac{2\pi m\,n}{b}\right)\right];  
\label{zeta3}
\end{equation}
\end{widetext}
using explicit formulas for $K_{\pm \frac{1}{2}}(z)=\sqrt{\frac{\pi}{2z}}e^{-z}$ and $\sum_{n=1}^{\infty}\frac{e^{-a}}{n}=-\ln\left(1-e^{-a}\right)$, 
we get from Eq.~(\ref{zeta1}) after some manipulations, 
 remembering $b=2\pi/L$, 
\begin{widetext}
\begin{equation}
\Pi(k^2,\beta)-\Pi(0,\beta)=  \ln
\left[\frac{\cosh\left(mL\sqrt{1+\frac{\beta(1-\beta)k^2}{m^2}} \right)-\cos
\theta}{\cosh\left( mL\right)-\cos \theta} \right].
\label{zeta3}
\end{equation}
\end{widetext}
This leads directly to the result
\begin{eqnarray}\label{eq:piren1}
\Pi^e_R (k^2) &=& \Pi_e (k^2) \,-\, \Pi_e(0) \nonumber\\
&=& \frac{2\,e^2}{\pi} \int_0^1 d\beta\; \beta(1-\beta)
\; \ln [ 1 + F(k^2) ]\;,\nonumber \\
\label{zeta4}
\end{eqnarray}
with
\begin{equation}\label{eq:piren2}
F(k^2) \;=\; \frac{\cosh \big[m L
\sqrt{1+\beta(1-\beta)\frac{k^2}{m^2}}\big] -
\cosh(m  L)}{\cosh (m L) - \cos(\theta)}\;.
\label{zeta5}
\end{equation}

It is interesting to note that, even though the theory is $5$ dimensional,
the vacuum polarization tensor requires, to be renormalized, just fixing the position 
and residue of one pole, as in $4$ dimensions. Indeed, the superficial
degree of divergence, $\delta (\gamma)$, for an $1PI$ Feynman graph
$\gamma$ in $QED_5$ is 
\begin{equation}
\delta(\gamma)= 5 - \frac{3}{2} E_G - 2 E_F + \frac{1}{2} V
\end{equation}
where $E_G$ and $E_F$ are the number of external gauge and fermion lines,
respectively, and $V$ is the number of vertices.
For the one-loop vacuum polarization tensor, we then have $\delta(\gamma)=
3$, which, taking into account gauge invariance is reduced to $1$.
Moreover, since the divergent terms can only be {\em even\/} polynomials in the
momentum, we are left with a zero degree divergence: this is the logarithmic divergence
already tamed in (\ref{eq:tamed}).

Let us now study some immediate properties and consequences that follow
from expressions (\ref{eq:piren1}) and (\ref{eq:piren2}) above. The natural
approach is perhaps to look at its predictions for different momentum
regimes. Let us thus begin by considering the low momentum regime, namely,
$k^2 << m^2$. The leading term, $k^2/m^2 \to 0$ has already been
considered, to impose the renormalization condition $\Pi_R^e \to 0$, which
is not actually a prediction, but rather the fact that the model contains
Coulomb's law at long distances.

The next-to-leading term already contains a non trivial effect. Indeed,  
a simple effect that will be sensible to the presence of the flux can be 
seen by expanding the renormalized tensor to  
 $(\frac{k}{m})^2$ order in a momentum expansion:
\begin{equation}
\Pi_R(k^2) \;\sim \; -\frac{e^2}{30 \pi} \Big[\frac{m L \,\,\sinh(m
L)}{\cos(\theta)-\cosh(m L)} \big]\, \frac{k^2}{m^2} \;,\; \; k^2 \sim 0 \,.
\end{equation}
The corresponding modification in the photon's effective action produces, for example,
a correction in the electrostatic potential due to a point charge. For the Hydrogen atom, the
corrected potential energy becomes:
\begin{equation}
V_{eff}(r) 
\;=\; - \frac{e^2}{4 \pi r} \,-\,\frac{e^4}{120 \pi^2 m^2} 
\Big[\frac{m L \,\sinh(m L)}{\cosh(m L) - \cos\theta}\Big] \; 
\delta^{(3)}({\mathbf r})\;. 
\end{equation}
The usual correction is obtained when $\theta \to 0$ and $m L \to 0$:
\begin{equation}
V_{eff} (r) \;\to\; - \frac{e^2}{4 \pi r} \,-\,\frac{e^4}{60 \pi^2 m^2} \; 
\delta^{(3)}({\mathbf r}) \;.
\end{equation}
It is interesting to study the shape of the ratio between the corrected and
usual strengths of the respective terms:
\begin{equation}
\xi(m L, \theta) \;\equiv\;\frac{ \frac{m L} {2} \,\sinh(m L)}{\cosh(m L) - \cos \theta}\;.  
\label{U1}
\end{equation}
The case of a vanishing flux yields simply 
$$\xi(m L, 0) \,=\,\frac{\frac{m L}{2}}{\tanh(\frac{m L}{2})},$$  
which for small values of $m L$ approaches $1$, and grows linearly with $m
L$ when \mbox{$m L >> 1$}. 

The opposite regime, when the effect of the flux is maximum, corresponds to
$\theta = \pi/2$:
\begin{equation}
\xi(m L, \frac{\pi}{2}) \;=\; \frac{m L} {2} \, \tanh(m L)\;.  
\label{U2}
\end{equation}
The behaviour in this case is quite different; it tends to zero
quadratically for small $m L$, and also grows linearly in the opposite
case, albeit with a different slope.

%%%%
It is noteworthy that, from eq.(\ref{U1}) one can obtain a crude
estimate for the length $L$.  To that end, we need some assumptions,  we
consider that we are within the vanishing flux aproximation, also as the typical
contribution of the vacuum polarization term for the energy shift in muonic
atoms is of the order of $0.5\,\%$ \cite{eides} we may then take
$\xi(m L, \,0) \lesssim 1.0001$; such a choice implies that a correction due to
an extra dimension does not  significantly changes the values from present data.
Having this in mind, we obtain through this simple reasoning that $L\,\lesssim\,
0.03 \,[m]^{-1}$ which in natural units can be translated to $L \lesssim
10^{-14}~\text{meters}$. 
In order to get a more stringent bound, one should take into account
other effects, which may show up different dependences on the physics of the
extra dimensions. That investigation is, however, outside the scope of the
present paper.
%%%%

Let us now consider the would be large-momentum region for the vacuum
polarization. This regime will be defined by the condition that \mbox{$k^2 >>
m^2$}, although $k$ (and $m$), will be assumed to be much smaller than
$L^{-1}$. The latter is enforced in order to say that the mass of the
Kaluza-Klein modes is much larger than the photon momentum. Under this
assumption, one gets the expression:
\begin{equation}
\Pi^e_R(k^2) \;\sim\; \frac{2\,e^2}{\pi} \int_0^1 d\beta\; \beta(1-\beta)
\, \ln \Big[ 1 +  \beta(1-\beta)\frac{k^2}{m_{eff}^2} \Big] \;,
\end{equation}
where
\begin{equation}
m_{eff} \;\equiv\; \frac{ 2 | \sin \frac{\theta}{2} | }{L} \;.
\end{equation}
We conclude that, as a consequence of the existence of the non-vanishing
flux, the large-momentum behaviour differs from the one that one has in
standard $QED$, by the emergence of an effective mass $m_{eff}$. This mass
should, in order not to spoil the known anti-screening effect at short
distances, be very small. Since $L$ is assumed to be very small, that can
only be achieved with an extremely small $\theta$, namely $\theta << 1$.
Hence,
\begin{equation}
m_{eff} \;\equiv\; \frac{ 2 | \sin \frac{\theta}{2} | }{L} \;\sim\;
\frac{|\theta|}{L} \; << \frac{1}{L} \;.
\end{equation}
In natural units, if $L^{-1} \equiv \Lambda$ is the large momentum scale set
by the Kaluza-Klein modes, and we want $m_{eff}$ to be much smaller than
the electron mass, since only in that situation we recover the expected
behaviour for the effective charge at small distances. Then we should have:
\begin{equation}
|\theta| < < \frac{m}{\Lambda}\;.
\end{equation}

\section{Parity breaking term}\label{sec:parity}

The parity breaking term, $\Gamma_o$ is simply obtained by taking into account
(\ref{eq:sumj}) and (\ref{eq:defj}):
\begin{equation}
\Gamma_o \,=\, - \frac{i e^2}{16 \pi^2} \,
\Phi \int d^{3+1}x {\tilde F}_{\mu\nu} F_{\mu\nu} \;,
\end{equation}
where we introduced the factor:
\begin{equation}
\Phi \,=\, \sum_{n=-\infty}^\infty \,\varphi_n \;;
\end{equation}
the sum of this series is well-known~\cite{Fosco:1997ei}, 
the result being:
\begin{equation}
\Phi \,=\, \arctan \big[ \tanh( \frac{m L}{2} ) \, \tan (\theta/2) \big] \;.  
\end{equation}
The possible effects due to this term are more difficult to elucidate,
since they would require the existence of non-trivial Abelian gauge field
background to manifest themselves. Within the present model, there is
no room to accommodate them, except if singular configurations were
included by hand.

\section{Conclusions}\label{sec:conc}

To conclude, we summarize the main points we have explored this paper:  
The vacuum polarization function exhibits physical effects due
to the extra dimension and flux.  Among those, the strongest one is due to the
non-vanishing flux, parametrized by $\theta$, and manifests itself in the large 
momentum behaviour of the effective charge. Indeed, $\theta$ should be much
smaller than the ratio between the electron mass and the (momentum) scale
induced by the inverse of the compactification radius for this
effect to be supressed.
Besides, the effect of the non-vanishing flux is maximum when it reaches
$\pi$. This is to be expected, since in that case there is no massless
mode, and hence there is no natural way to dimensionally reduce 
the theory at the level of the fermionic field. That is, on the other hand, 
the case when $\theta=0$, since it means that the $n=0$ mode finds a natural 
$3+1$ dimensional interpretation and there is a smooth limit when $L \to 0$. 
We find that 
Finally, parity breaking effects might be expected only if there were
a compelling reason to know that the gauge field itself adopts a
topologically non-trivial configuration; this cannot be done
within the context of the present model.  

\section*{Acknowledgements}
C.D.F. and C.C.T. thank CONICET for financial
support.  A.P.C.M.  and I.R. thank CNPq/MCT and FAPERJ
for partial financial support.

%%%%%%%%%%%%%%%%%%%%%%%%%%%%%%%%%%%%%%%%%%%%%%%%%%%%%%%%%%%%%%%%%%%%%%%%%
%%%%%%%%%%%%%%%%%%%%%%%%%%% References %%%%%%%%%%%%%%%%%%%%%%%%%%%%%%%%%%
%%%%%%%%%%%%%%%%%%%%%%%%%%%%%%%%%%%%%%%%%%%%%%%%%%%%%%%%%%%%%%%%%%%%%%%%%

\end{document}